\documentclass[prb,twocolumn,amsmath,amssymb]{revtex4}
\usepackage{graphicx}% Include figure files
\usepackage{dcolumn}% Align table columns on decimal point
\usepackage{bm}% bold math

\begin{document}

%\preprint{APS/123-QED}

\title{Lifetime of Single-Particle Excitations in a Dilute Bose-Einstein Condensate \\at Zero Temperature}

\author{Kazumasa T{\sc sutsui} and Takafumi K{\sc ita}}
\affiliation{Department of Physics, Hokkaido University, Sapporo 060-0810, Japan}%

\date{\today}

\begin{abstract}
We study the lifetime of single-particle excitations in a dilute homogeneous Bose-Einstein condensate at zero temperature based on 
a self-consistent perturbation expansion of satisfying Goldstone's theorem and conservation laws simultaneously.
It is shown that every excitation for each momentum ${\bm p}$ should have a finite lifetime proportional to the inverse $a^{-1}$ of  the $s$-wave scattering length $a$,
instead of $a^{-2}$ for the normal state, 
due to a new class of Feynman diagrams for the self-energy that emerges upon condensation.
We calculate the lifetime as a function of $|{\bm p}|$ approximately.
\end{abstract}

%\pacs{03.70.+k, 03.75.Hh, 03.75.Kk, 67.85.-d}% PACS, the Physics and Astronomy
\maketitle

The interaction between particles
yields a finite decay rate in every single-particle excitation of many-particle systems.
It is caused by collisions between particles that are describable as second- and higher-order processes of the perturbation expansion
in terms of the interaction.\cite{AGD63}
Hence, one may expect generally that the decay rate  $\Gamma$, which is the inverse of the lifetime $\tau$ ($\hbar=1$), 
depends quadratically for a dilute system on the $s$-wave scattering length $a$
as $\Gamma\propto a^2$.
We will show below, however, that this is not the case for Bose-Einstein condensates (BECs) 
where $\Gamma$ will be proportional to $a$.

Theoretical attempts to microscopically describe Bose-Einstein condensates have encountered fundamental difficulties 
due to a finite thermodynamic average $\langle \psi\rangle$ of the field operator $\psi$ itself,
such as the conserving-gapless dilemma\cite{HM65,Griffin96} and infrared divergences.\cite{GN64}
To resolve them, a self-consistent perturbation expansion has been constructed recently
in such a way as to satisfy a couple of exact statements simultaneously, i.e., conservation laws and Goldstone's theorem.\cite{Kita09}
According to it,\cite{Kita09,Kita11} there should be a new type of Feynman diagrams for the self-energy
that are classified as ``one-particle reducible'' (1PR) or ``improper'' in the normal state,\cite{AGD63,LW}
which may modify standard results based on the Bogoliubov theory\cite{AGD63,Bo} substantially.
For example, we have predicted in a previous paper that they will convert the Lee-Huang-Yang expressions\cite{LHY57} 
for the ground-state energy per particle $E/N$ and condensate density $n_0$ of the dilute Bose gas into\cite{TK13}
\begin{subequations}
\label{En_0}
\begin{align}
&\frac{E}{N}=\frac{2\pi\hbar^2 an}{m}\left[\,1+\left(\frac{128}{15\sqrt{\pi}}+\frac{16}{5}c_{\rm ip}\right)\!\sqrt{a^3n}\,\right],\\
&\frac{n_0}{n}=1-\left(\frac{8}{3\sqrt{\pi}}+c_{\rm ip}\right)\!\sqrt{a^3n},
\end{align}
\end{subequations}
where $m$ and $n$ are the particle mass and density, respectively, 
and $c_{\rm ip}$ is an extra constant of order $1$ due to those diagrams.

In the present paper, we focus on the lifetime 
of single-particle excitations in a dilute BEC at zero temperature.
We predict that the 1PR diagrams, which are characteristic of BECs, transform the nature of the Bogoliubov mode substantially into
a ``bubbling'' mode\cite{Kita11} with a proper lifetime proportional to $a^{-1}$.

We consider a homogeneous system of identical Bose particles with mass $m$ and spin $0$ interacting via the contact
potential $U\delta({\bm r}-{\bm r}')$. 
The Hamiltonian is given by
\begin{align}
H =\sum_{\bm p}(\epsilon_{p}\!-\!\mu)c^\dagger_{\bm p}c_{\bm p}+\frac{U}{2V}\!\sum_{{\bm p}{\bm p}'{\bm q}}c^\dagger_{{\bm p}+{\bm q}}c^\dagger_{{\bm p}'-{\bm q}}c_{{\bm p}'}c_{{\bm p}},
\end{align}
where ${\bm p}$, $\epsilon_{p}\!\equiv\!p^2/2m$, $\mu$, and $V$ are the momentum, kinetic energy, 
chemical potential, and volume, and $c^{\dagger}_{\bm p}$ and $c_{\bm p}$ are the creation and annihilation operators, respectively.
We set $\hbar=2m=k_{\rm B}=T_{\rm c}^{0}=1$,
where $k_{\rm B}$ denotes the Boltzmann constant and $T_{\rm c}^{0}$ is the transition temperature of ideal Bose-Einstein condensation.\cite{FW71}
Ultraviolet divergences inherent in the continuum model are removed here by introducing a momentum cutoff $p_c\gg 1$. 
It is standard in the low-density limit to express $U$ in terms of the s-wave scattering length $a$.
They are connected in the conventional units by
\begin{eqnarray*}
\frac{m}{4\pi\hbar^2 a}=\frac{1}{U}+\int\frac{d^3p}{(2\pi\hbar)^3}\frac{\theta(p_c-p)}{2\epsilon_{p}},
\end{eqnarray*}
with $\theta(x)$ the step function, which in the present units reads 
\begin{align}
1/8\pi a=1/U+p_c/4\pi^2.
\label{U-a}
\end{align}
We will focus on the limit $a\rightarrow 0$ and choose $p_c$ so that $1\ll p_c\ll \pi/2a$ is satisfied;
thus, $U=8\pi a$ to the leading order.

Green's function for a homogeneous BEC can be expressed in the Nambu representation as\cite{Kita09,Kita11}
\begin{subequations}
\label{hatG}
\begin{align}
\hat G_{\vec p}=\begin{bmatrix}{G_{\vec{p}}} &{F_{\vec{p}}} \vspace{2mm}\\{-\bar F_{\vec{p}}} &{-\bar G_{\vec{p}}}\end{bmatrix},
\label{hatG1}
\end{align}
where $\vec{p}=({\bm p},z_{\ell})$ with $z_\ell\equiv 2\pi T\ell\, i \ (\ell=0,\pm1,\pm,2\cdots)$ and $T$ the temperature. 
The upper elements satisfy $G_{\vec p}\!=\!G^{*}_{{\vec p}^*}$  and $F_{\vec p}\!=\!F_{-\vec{p}}$, and a barred quantity generally denotes 
$\bar{G}_{\vec p}\!=\!G^*_{-\vec{p}^*}$.\cite{Kita11}
This matrix $\hat G_{\vec p}$ obeys the Dyson-Beliaev equation\cite{Beliaev58,Kita09}
\begin{align}
\hat G_{\vec p}&=\begin{bmatrix}{z_{\ell}-\epsilon_{p}-\Sigma_{\vec p}+\mu} &{-\Delta_{\vec p}} \vspace{2mm}\\{\bar\Delta_{\vec p}} &{z_{\ell}+\epsilon_{p}+\bar\Sigma_{\vec p}-\mu}\end{bmatrix}^{-1}
\label{DBe},
\end{align}
\end{subequations}
which may also be regarded as defining the self-energies $\Sigma_{\vec p}$ and $\Delta_{\vec p}$\hspace{0.2mm}.
In the self-consistent perturbation expansion, they are obtained from a functional $\Phi[\hat G_{\vec p}, n_0]$ as\cite{Kita09,Kita11}
\begin{subequations}
\label{Phi-Sigma}
\begin{align}
\Sigma_{\vec{p}}=-\frac{1}{T}\frac{\delta \Phi}{\delta G_{\vec{p}}},\hspace{10mm} \Delta_{\vec{p}}=\frac{2}{T}\frac{\delta \Phi}{\delta \bar F_{\vec{p}}}.
\label{Phi-Sigma1}
\end{align}
In addition, $\Phi$ satisfies
\begin{align}
\frac{1}{V}\frac{\delta \Phi}{\delta n_0}=\Sigma_{\vec 0}-\Delta_{\vec 0}
\label{Phi-Sigma2}
\end{align}
\end{subequations}
in a gauge where $\Delta_{\vec 0}$ is real.
Substitution of Eq.\ (\ref{Phi-Sigma1}) into Eq.\ (\ref{DBe}) yields
self-consistent (i.e., nonlinear) equations for $G_{\vec{p}}$ and $F_{\vec{p}}$.
It also follows from Eq.\ (\ref{Phi-Sigma2}) that the extremal condition $\delta \Omega/\delta n_0\!=\!0$ for the thermodynamic potential $\Omega$ yields
the Hugenholtz-Pines relation\cite{HP59}
\begin{align}
\mu=\Sigma_{\vec 0}-\Delta_{\vec 0}.
\label{HPr}
\end{align}
Equations (\ref{hatG})-(\ref{HPr}) are exact statements.
It has been shown that the functional $\Phi$ can be constructed as a power-series expansion in $U$ in such a way that Eq.\ (\ref{Phi-Sigma}), 
conservation laws (i.e., Noether's theorem), and an exact relation for the interaction energy are satisfied simultaneously order by order.\cite{Kita09}

\begin{figure}[t]
\begin{center}
\includegraphics[width=0.75\linewidth]{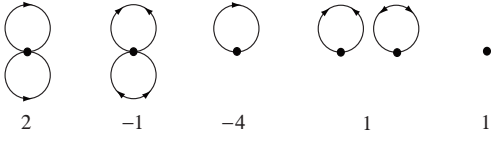}
\end{center}
\vspace{-4mm}
\caption{Feynman diagrams for $\Phi^{(1)}$. 
A filled circle denotes $U$, a line with an arrow (two arrows) represents $G$
(either $F$ or $\bar{F}$) in Eq.\ (\ref{hatG1}) as in the theory of superconductivity,\cite{AGD63,FW71}
and every missing line in the last three diagrams corresponds to $n_0$.
The number below each diagram indicates its relative weight,
which should be multiplied by $V/2$ to obtain the absolute weight.\label{Fig1}}
\end{figure}

Let us write down the key functional $\Phi$ perturbatively.\cite{Kita09,Kita11}
The first-order terms are given diagrammatically in Fig.\ \ref{Fig1},
which analytically reads 
\begin{align}
\Phi^{(1)}=&\,\frac{V}{2}U \left[\sum_{\vec{p}_1 \vec{p}_2}\bigl(2G_{\vec{p}_1}G_{\vec{p}_2}-F_{\vec{p}_1}\bar F_{\vec{p}_2}\bigr)\right.
\nonumber \\
& \left.+\,n_0 \sum_{\vec p}\bigl(-4 G_{\vec p}+F_{\vec{p}}+\bar F_{\vec{p}}\bigr)+n_0^2\right] ,
\label{Phi^(1)}
\end{align}
with the summation over $\vec{p}$ defined by
$$
\sum_{\vec{p}}\equiv T\sum_{\ell=-\infty}^{\infty}\frac{1}{V}\sum_{\bm p} .
$$
It should be noted that 
the relative importance of each term in Eq.\ (\ref{Phi^(1)}) for the dilute limit at $T\!=\!0$ increases with the power of $n_0$.

\begin{figure}[t]
\begin{center}
\includegraphics[width=0.8\linewidth]{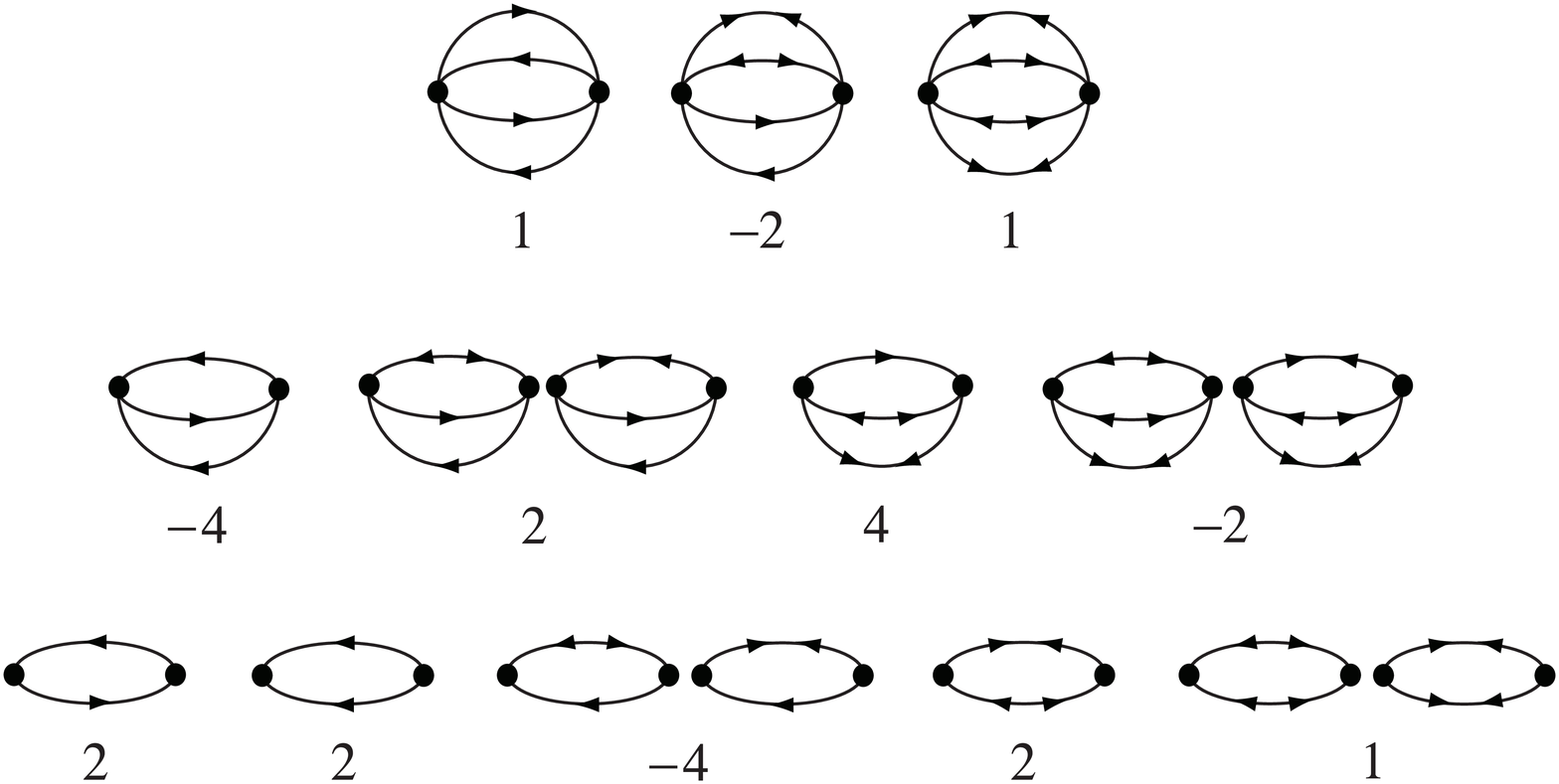}
\end{center}
\vspace{-4mm}
\caption{Feynman diagrams for $\Phi^{(2)}$. 
The number below each diagram indicates its relative weight,
which should be multiplied by $-V/2$ to obtain the absolute weight.\label{Fig2}}
\end{figure}

Next, Fig.\ \ref{Fig2} enumerates second-order diagrams.
Dominant among them in the dilute limit at $T\!=\!0$ are those
in the third row proportional to $n_0^2$, i.e., those with the highest power in terms of $n_0$.
Their contribution can be expressed concisely as
\begin{subequations}
\label{Phi^23ip}
\begin{equation}
\Phi^{(2{\rm ip})}=-\frac{V}{2}(Un_0)^2\sum_{\vec p}(G_{\vec p}+\bar G_{\vec p}-F_{\vec p}-\bar F_{\vec p})^2 ,
\label{Phi^2ip}
\end{equation}
where $2G_{\vec p}\bar G_{\vec p}$, for example, corresponds to the first diagram in the third row of Fig.\ \ref{Fig2},
whereas  both $G_{\vec p}G_{\vec p}$ and $\bar G_{\vec p}\bar G_{\vec p}$ are associated with the second particle-particle bubble diagram 
to yield the same contribution.

\begin{figure}[b]
\begin{center}
\includegraphics[width=0.8\linewidth]{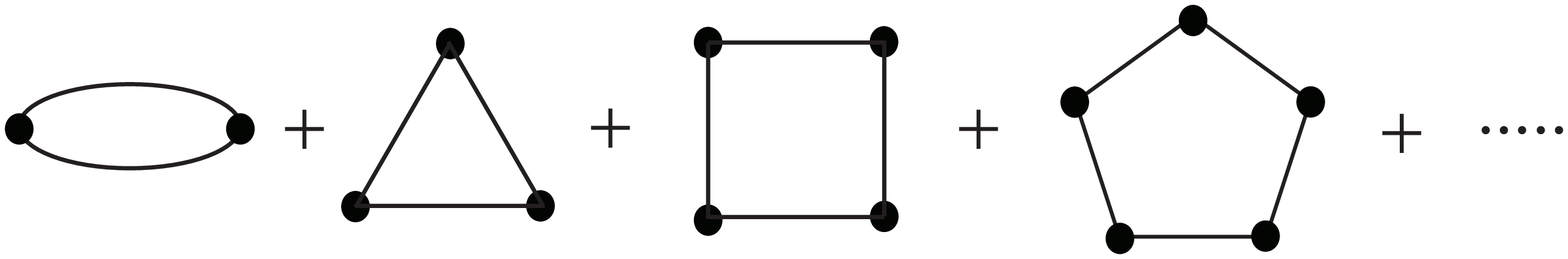}
\end{center}
\vspace{-4mm}
\caption{Feynman diagrams for $\Phi$ beyond the first order that are dominant in the dilute limit at $T\!=\! 0$.
Arrows are suppressed.\label{Fig3}}
\end{figure}

Extending the analysis to higher orders, one may be convinced that the leading contribution beyond the first order originates 
from the series of Fig.\ \ref{Fig3}.
These diagrams are characteristic of BECs to produce unusual 1PR self-energies 
upon the differentiations of Eq.\ (\ref{Phi-Sigma1}).
However, they result naturally from the requirement that Goldstone's theorem be satisfied order by order in $U$.\cite{Kita09,Kita11}
They are responsible for the constant $c_{\rm ip}$ in Eq.\ (\ref{En_0}),\cite{TK13}
and also bring about a finite lifetime proportional to $a^{-1}$ in the single-particle excitations, as shown below.
To be specific, the third-order contribution can be written as\cite{Kita09,TK13}
\begin{equation}
\Phi^{(3{\rm ip})}=-\frac{5V}{3}(Un_0)^3\sum_{\vec p}(G_{\vec p}+\bar G_{\vec p}-F_{\vec p}-\bar F_{\vec p})^3 ,
\label{Phi^3ip}
\end{equation}
\end{subequations}
where factor $5$ is understood as a sum of $1$ and $2^2$
originating from the weights of the normal particle-particle and particle-hole bubble diagrams, respectively.\cite{TK13}
Both Eqs.\ (\ref{Phi^2ip}) and (\ref{Phi^3ip}) are given solely as a functional of
\begin{align}
f_{\vec p}\equiv G_{\vec p}+\bar G_{\vec p}-F_{\vec p}-\bar F_{\vec p} 
\label{f_p}
\end{align}
that satisfies $f_{\vec p}=\bar f_{\vec p}$. 
The statement also holds true for higher-order diagrams in the series of Fig.\ \ref{Fig3}.

Thus, our approximate $\Phi$'s adopted below are all expressible as
\begin{equation}
\Phi[\hat G_{\vec p},n_0]\approx \Phi^{(1)}[\hat G_{\vec p},n_0]+\Phi^{({\rm ip})}[f_{\vec p},n_0],
\label{Phi-approx}
\end{equation}
where $\Phi^{(1)}$ is given by Eq.\ (\ref{Phi^(1)}), and $\Phi^{({\rm ip})}[f_{\vec p},n_0]$ denotes some partial contribution 
from the infinite series of Fig.\ \ref{Fig3}.
To be specific, we consider the three approximations
\begin{subequations}
\label{Phi^(ip)}
\begin{align}
&\Phi^{({\rm ip})}\approx -V\sum_{\vec p} \frac{1}{2}\bigl(Un_0f_{\vec p}\bigr)^2 , \\
&\Phi^{({\rm ip})}\approx -V\sum_{\vec p} \!\left[\frac{1}{2}\bigl(Un_0f_{\vec p}\bigr)^2+\frac{5}{3}\bigl(Un_0f_{\vec p}\bigr)^3\right]\! , \\
&\Phi^{({\rm ip})}\approx -V\sum_{\vec p} \!\left[ \frac{1}{2}\bigl(Un_0f_{\vec p}\bigr)^2+\sum_{n=3}^\infty \frac{1+2^{n-1}}{n}(Un_0 f_{\vec p})^{n}   \right]\! .
\end{align}
\end{subequations}
The first two correspond to $\Phi^{({\rm ip})}\!\approx\! \Phi^{(2{\rm ip})}$ and $\Phi^{({\rm ip})}\!\approx\! \Phi^{(2{\rm ip})}\!+\!\Phi^{(3{\rm ip})}$ from Eq.\ (\ref{Phi^23ip}),
respectively, whereas the last one incorporates the contribution originating from the particle-particle and particle-hole bubbles 
up to the infinite order. We call them as the second-order, third-order, and fluctuation-exchange (FLEX) approximations, respectively.

The self-energies are obtained subsequently by inserting Eq.\ (\ref{Phi-approx}) into Eq.\ (\ref{Phi-Sigma1}),
whose differentiations graphically correspond to removing a line of 
$G_{\vec p}$ and $\bar F_{\vec p}$ from every diagram for $\Phi$ in all possible ways, respectively.
Note also  that  $G_{\vec p}$ and $\bar G_{\vec p}$ in Eq.\ (\ref{f_p}) yield
the same contribution upon the differentiation in terms of $G_{\vec p}$. 
It follows from Eq.\ (\ref{Phi^(1)}) and $\Phi^{({\rm ip})}\!=\!\Phi^{({\rm ip})}[f_{\vec p},n_0]$ in Eq.\ (\ref{Phi-approx}) 
that the self-energies in this approximation can be expressed as
\begin{subequations}
\label{SD}
\begin{align}
\Sigma_{\vec p}=&\,\,\Sigma^{(1)}+\Delta_{\vec p}^{({\rm ip})},
\label{SD1}
\\
\Delta_{\vec p}=&\,\,\Delta^{(1)}+\Delta_{\vec p}^{({\rm ip})} .
\label{SD2}
\end{align}
\end{subequations}
Here, the first-order self-energies are given by
\begin{subequations}
\label{D1n1}
\begin{align}
\Delta^{(1)}= U\left(n_0-\sum_{\vec p} F_{\vec p}\right) ,
\label{D1}
\end{align}
and $\Sigma^{(1)}= 2Un$ with 
\begin{align}
n\equiv n_0-\sum_{\vec p}G_{\vec p}\ e^{z_{\ell}0_+}
\label{n1}
\end{align}
\end{subequations}
denoting the particle density and $0_+$ an infinitesimal positive constant.
It is worth pointing out that $n={\zeta(3/2)}/{(4\pi)^{3/2}}$ in the present units with $\zeta(3/2)\!=\!2.612\cdots$ the Riemann zeta function.\cite{TK12}
Next, $\Delta_{\vec p}^{({\rm ip})}$ is obtained for each approximate $\Phi^{({\rm ip})}$ in Eq.\ (\ref{Phi^(ip)}) 
as
\begin{subequations}
\label{TD}
\begin{align}
\Delta^{\rm (ip)}_{\vec{p}}=& \,\,2(Un_0)^2f_{\vec p} \, , \label{TD2} \\
\Delta^{\rm (ip)}_{\vec{p}}=& \,\,2(Un_0)^2f_{\vec p}+10(Un_0)^3f_{\vec p}^2\,  , \label{TD3} \\
\Delta^{\rm (ip)}_{\vec{p}}=&\,\,\frac{(2Un_0)^2f_{\vec{p}}}{1-2Un_0f_{\vec{p}}}+\frac{2(Un_0)^2f_{\vec{p}}}{1-Un_0f_{\vec{p}}}-4(Un_0)^2f_{\vec{p}}  \,\label{TDF} ,
\end{align}
\end{subequations}
respectively.
It follows from $f_{\vec p}=\bar f_{\vec p}$ in Eq.\ (\ref{f_p}) that $\Delta^{\rm (ip)}_{\vec{p}}$ also satisfies
$\Delta^{\rm (ip)}_{\vec{p}}=\bar \Delta^{\rm (ip)}_{\vec{p}}$.

Now that we have written down the self-energies explicitly, 
we substitute Eq.\ (\ref{SD}) into Eq.\ (\ref{HPr}). We then find the Hugenholtz-Pines relation in our approximation as
\begin{align}
\mu=\Sigma^{(1)}-\Delta^{(1)}.
\label{mu1}
\end{align}
Next, let us substitute Eqs.\ (\ref{SD}) and (\ref{mu1}) into Eq.\ (\ref{DBe}) and perform the matrix inversion.
We thereby obtain
\begin{align}
\hat G_{\vec p}=\frac{1}{z_\ell^2-\epsilon_{p}(\epsilon_{p}+2\Delta_{\vec p})}\!
\begin{bmatrix}
z_\ell+\epsilon_{p}+\Delta_{\vec p} \!\!&\!\! \Delta_{\vec p} \vspace{2mm} \\
-\Delta_{\vec p} \!\!&\!\! z_\ell-\epsilon_{p}-\Delta_{\vec p} 
\end{bmatrix},
\label{G1}
\end{align}
so that Eq.\ (\ref{f_p}) is expressed concisely as
\begin{align}
f_{\vec p}= \frac{2\epsilon_{p}}{z_\ell^2-\epsilon_{p}(\epsilon_{p}+2\Delta_{\vec p})}.
\label{f_p-2}
\end{align}
The spectral function, which has full information on the single-particle excitations,
is obtained from Eq.\ (\ref{G1}) by\cite{AGD63,FW71}
\begin{align}
A_{p}(\omega)=-2\,{\rm Im}\left.G_{\vec{p}}\ \right|_{z_\ell\rightarrow\omega+i0_+}.
\label{A_p}
\end{align}
This completes our formulation.

Now, our numerical procedure to calculate the spectral function is summarized as follows. 
Green's function (\ref{G1}) is given as a functional of  $(z_\ell,\epsilon_{p},\Delta_{\vec p})$
with $\Delta_{\vec p}\!=\!\Delta^{(1)}\!+\! \Delta_{\vec p}^{({\rm ip})}$, 
where $\Delta_{\vec p}^{({\rm ip})}$ is determined as a solution of the algebraic equation (\ref{TD}) with Eq.\ (\ref{f_p-2})
for a given set of $(z_\ell,\epsilon_{p},\Delta^{(1)},Un_0)$.
Besides, it follows from Eqs.\ (\ref{D1n1}) and (\ref{U-a}) that
\begin{equation}
\Delta^{(1)}\approx Un_0 \approx 8\pi n a=\frac{\zeta(3/2)}{\sqrt{\pi}}a
\label{D^(1)-un_0}
\end{equation}
to the leading order in $a$.
Adopting this approximation, we can solve Eq.\ (\ref{TD}) with Eq.\ (\ref{f_p-2}) as a function of  $(z_\ell/\Delta^{(1)},\epsilon_{p}/\Delta^{(1)})$ 
so as to satisfy $\Delta_{\vec p}^{({\rm ip})}\rightarrow 0$ for $|z_{\ell}|\rightarrow \infty$.
The resultant values on the imaginary axis are used subsequently to 
calculate Eq.\ (\ref{A_p})
based on Thiele's reciprocal difference algorithm for Pad\'{e} approximants.\cite{Pade}
Function $A_{p}(\omega)$ thereby calculated can be checked numerically with a couple of exact relations
\begin{subequations}
\begin{align}
&\int_{-\infty}^{\infty}\frac{d\omega}{2\pi}A_{p}(\omega)=1,
\label{sum}
\\
&G_{\vec p}=\int_{-\infty}^{\infty}\frac{d\omega}{2\pi}\frac{A_{p}(\omega)}{z_\ell-\omega}.
\label{reb}
\end{align}
\end{subequations}
We have confirmed that sum rule (\ref{sum}) is satisfied beyond $99.7$\%, and
Green's function on the imaginary axis are reproduced with an error of less than $0.02$\%.

To start with, let us review the spectral function of the Bogoliubov theory,
which is obtained by inserting Eq.\ (\ref{G1}) with $\Delta_{\vec p}\rightarrow \Delta^{(1)}$ into Eq.\ (\ref{A_p}).
It reads
\begin{align}
A_{p}^{(1)}(\omega)=\pi\bigl[(\alpha_{p}+1)\delta\bigl(\omega-E^{\rm B}_{p}\bigr)-(\alpha_{p}-1)\delta\bigl(\omega+E^{\rm B}_{p}\bigr)\bigr],
\end{align} 
where $E_{p}^{\rm B}\!\equiv\!\sqrt{\epsilon_{p}(\epsilon_{p}\!+\! 2\Delta^{(1)})}$ is the Bogoliubov spectrum 
with a linear $p$ dependence for $\epsilon_p/\Delta^{(1)}\ll 1$, 
and $\alpha_{p}\equiv(\epsilon_{p}+\Delta^{(1)})/E_{p}^{\rm B}$.
Thus, $A_{p}^{(1)}(\omega)$ has a couple of sharp $\delta$-function peaks at $\omega=\pm E^{\rm B}_{p}$
corresponding to well-defined quasiparticles with the infinite lifetime.

\begin{figure}[t]
\begin{center}
\includegraphics[width=0.9\linewidth]{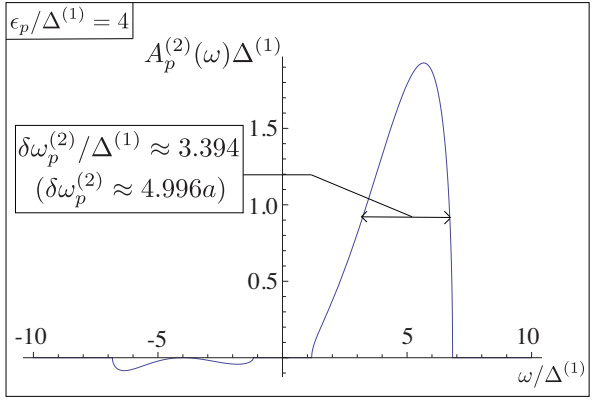}
\end{center}
\caption{Plot of the spectral function $A^{(2)}_{p}(\omega)$ in the second-order approximation for $\epsilon_{p}/\Delta^{(1)}=4$.   
The half width $\delta \omega^{(2)}_{p}$ is proportional to $a$.}\label{2nd_A4n}
\end{figure}
\begin{figure}[t]
        \begin{center}
                \includegraphics[width=0.9\linewidth]{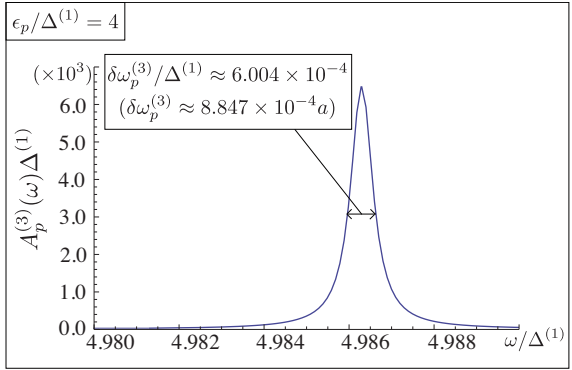}
               \end{center}
\caption{Plot of the spectral function $A^{(3)}_{p}(\omega)$  in the third-order approximation  for $\epsilon_{p}/\Delta^{(1)}=4$. \label{3rd_A4n}}
\end{figure}

However, the ``improper'' self-energy $\Delta^{({\rm ip})}_{\vec{p}}$ brings about a qualitative change in the spectral function.
To see this explicitly, let us consider the second-order approximation of Eq.\ (\ref{TD2}) with Eqs.\ (\ref{f_p-2}) and (\ref{D^(1)-un_0}),
which can be solved analytically as 
$\Delta^{({\rm ip})}_{\vec p}={D^{(1)}_{\vec p}}/{4\epsilon_{p}}-\bigl[\bigl({D^{(1)}_{\vec p}}/{4\epsilon_{p}}\bigr)^2-2\bigl(\Delta^{(1)}\bigr)^2\bigr]^{1/2}$
with $D^{(1)}_{\vec p}\equiv(z_\ell-E_{p}^{\rm B})(z_\ell+E_{p}^{\rm B})$.
Using this $\Delta^{({\rm ip})}_{\vec p}$ in Eq.\ (\ref{G1}) and substituting the resultant $G_{\vec p}$ into Eq.\ (\ref{A_p}), 
we obtain the spectral function in the second-order approximation as
\begin{align}
A^{(2)}_{p}(\omega)=&\,\,\theta\bigl(\omega^2-E_{p_-}^2\bigr)\theta\bigl(E_{p_+}^2-\omega^2\bigr)[\theta(\omega)-\theta(-\omega)]\notag\\ 
&\,\,\times\frac{(\omega+\epsilon_{p})^2}{(4\Delta^{(1)})^2\epsilon_{p}^3}\sqrt{(E_{p_+}^2-\omega^2)(\omega^2-E_{p_-}^2)},
\label{A^(2)}
\end{align}
with $E_{p_\pm}\!\equiv \!\sqrt{\epsilon_{p}[\epsilon_{p}\!+\! (2\!\pm\!4\sqrt{2})\Delta^{(1)}]}$. 
Figure \ref{2nd_A4n} exhibits $A^{(2)}_{p}(\omega)$ for $\epsilon_{p}/\Delta^{(1)}=4$. 
As seen clearly, the quasiparticle peaks are broadened substantially due to $\Delta^{({\rm ip})}_{\vec p}$.
The half width  $\delta \omega^{(2)}_{p}\propto \tau_p^{-1}$  of the main peak for $\omega>0$ is clearly of the order of $a$ and 
approaches  $7.21a$ as $\epsilon_{p}\rightarrow\infty$.
However, Eq.\ (\ref{A^(2)}) is valid only for $\epsilon_{p}/\Delta^{(1)}\ge-2+4\sqrt{2}$; the second-order approximation
fails to describe the low-momentum region adequately.

\begin{figure}[t]
        \begin{center}
                \includegraphics[width=0.9\linewidth]{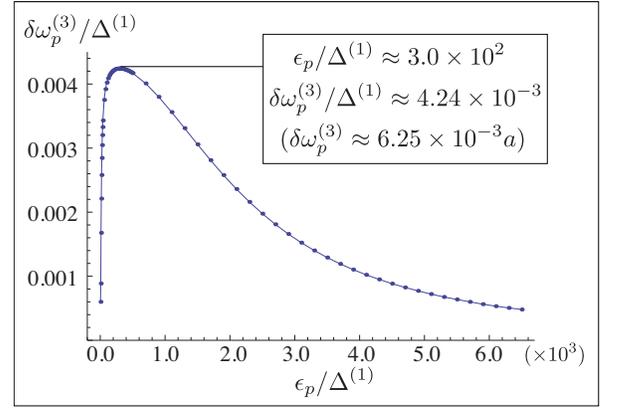}
                \end{center}
\caption{Half width $\delta \omega^{(3)}_{p}$ in the third-order approximation as a function of $\epsilon_{p}/\Delta^{(1)}$.\label{3rd_delAn}}
\end{figure}
\begin{figure}[t]
        \begin{center}
                \includegraphics[width=0.9\linewidth]{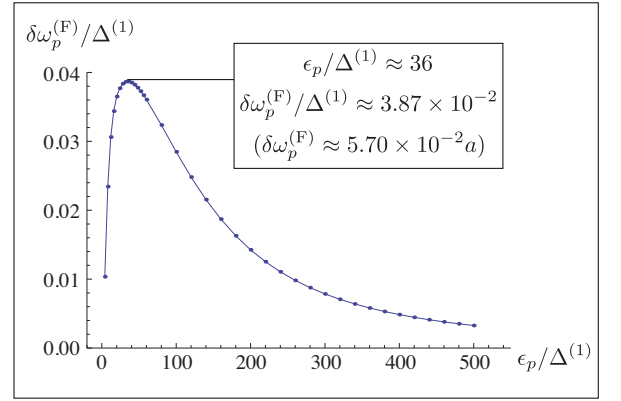}
                \end{center}
\caption{Half width $\delta\omega^{({\rm F})}_{p}$ in the FLEX approximation as a function of $\epsilon_{p}/\Delta^{(1)}$.\label{FLEX_delAn}}
\end{figure}

This unphysical behavior is removed in the third-order approximation of solving Eq.\ (\ref{TD3}) with Eqs.\ (\ref{f_p-2}) and (\ref{D^(1)-un_0}).
Figure \ref{3rd_A4n} plots the spectral function $A^{(3)}_{p}(\omega)$ in the third-order approximation for $\epsilon_{p}/\Delta^{(1)}=4$
around its peak for $\omega>0$. 
As seen clearly, this peak also has a finite width proportional to $a$, implying a finite lifetime $\tau_p\propto a^{-1}$
in the quasiparticle excitation.
Figure \ref{3rd_delAn} shows $\epsilon_{p}$ dependence of $\delta\omega^{(3)}_{p}\propto \tau_p^{-1}$.
It apparently develops from zero as $\epsilon_{p}$ is increased, has the maximum around $\epsilon_{p}/\Delta^{(1)}\approx 300$,
and starts to decrease thereafter towards zero.
A qualitatively similar behavior is obtained by the FLEX approximation 
of solving Eq.\ (\ref{TDF}) with Eqs.\ (\ref{f_p-2}) and (\ref{D^(1)-un_0}), as shown in Fig.\ \ref{FLEX_delAn}.
However, both the magnitude of $\delta\omega^{({\rm F})}_{p}$ and its peak location are quantitatively  different from $\delta\omega^{(3)}_{p}$.
To resolve this point requires a better treatment of the infinite series of Fig.\ \ref{Fig3}.

In summary, we have clarified that every single-particle excitation in diute BECs should have a proper lifetime even at $T\!=\!0$ 
that is proportional to the inverse of the $s$-wave scattering length $a$,
because of the 1PR diagrams for the self-energy.
The proportionality constant of the half-width $\delta\omega_p$ 
develops from zero at $p\!=\! 0$, increases as momentum $p$ gets larger to have a maximum, and expected to approach $0$  eventually for $p\!\rightarrow\!\infty$.

\end{document}